\documentclass[twocolumn]{aa}

\usepackage{graphicx,txfonts,natbib,verbatim}
\bibpunct{(}{)}{;}{a}{}{,}

\begin{document}

\title{Parallel electric field amplification by phase-mixing of Alfven waves}

\author{N.H. Bian, E. P. Kontar}

\offprints{N.H. Bian \email{nbian@astro.gla.ac.uk}}

\institute{Department of Physics \& Astronomy, University of Glasgow, G12 8QQ, United Kingdom}

\date{Received ; Accepted }

\abstract{Previous numerical studies have identified "phase mixing" of low-frequency Alfven waves as a mean of parallel
electric field amplification and acceleration of electrons in a collisionless plasma.}{Theoretical explanations
are given of how this
produces an amplification of the parallel electric field, and as a consequence, also leads to enhanced collisionless damping
of the wave by energy transfer to the electrons.}
{Our results are based on the properties of the Alfven waves in a warm plasma which are obtained from drift-kinetic theory, in particular, the rate of their electron Landau damping.}
{Phase mixing in a collisionless low-$\beta$ plasma proceeds in a manner very similar to the visco-resistive case, except for the
fact that electron Landau damping is the primary energy dissipation channel. The time and length scales
involved are evaluated. We also focus on the evolution of the parallel electric field and calculate its maximum
value in the course of its amplification}
{}


\titlerunning{Parallel electric field generation by kinetic Alfven wave turbulence}

\authorrunning{Bian et al}

\maketitle

\section{Introduction}
At finite wave-numbers in the direction perpendicular
to the ambient magnetic field, Alfven waves produce a compression of the plasma
which results in the creation of a parallel electric field via
the thermo-electric effect, i.e. due to electron pressure fluctuations along the magnetic field
lines. This electric field, whose magnitude increases with $k_{\perp}$, leads to wave-particle
interactions, and hence, to collisionless damping of the wave.

Importance of this
parallel electric field was pointed out already some time ago by \citep{Hasegawa1976a}.
Indeed, they argue that "resonant absorption" \citep{Hasegawa1974}
is a manifestation of mode conversion from the MHD Alfven wave (AW) to the kinetic Alfven wave (KAW)
and that the physical mechanism of the heating depends on the collisionless absorption
of the KAW. Although the original motivation was heating electrons
in laboratory fusion plasmas, this electric field was also proposed as a mechanism which
can accelerate electrons in space plasmas\citep{Hasegawa1976b,Hasegawa1978,Hasegawa1985,Goertz1979} and
for understanding solar coronal heating\citep{Ionson1978}.

\citep{Heyvaerts1983}
also introduced the idea of "phase-mixing" to improve the efficiency of AW dissipation.
Their theory is based on visco-resistive magnetohydrodynamics (MHD).
 Since then, MHD phase mixing has attracted
a significant amount of attention in the context of heating open magnetic structures in the solar corona\citep{Parker1991,
Nakariakov1997, Botha2000, DeMoortel2000, Hood2002}. Popular excitation mechanisms for coronal AWs in open magnetic structures
are photospheric motions and chromospheric reconnection events, respectively for the low-frequency and high-frequency 
range of the spectrum. 

Phase mixing can be understood
as the refraction of the wave while it propagates along a magnetic field with transverse
variation in the Alfven velocity, i.e. the progressive increase of its $k_{\perp}$. This is
a special occurrence of conservative energy cascade\citep{Bian2008}, a phenomenon 
generally attributed to non-linear interactions. Therefore, it is not
surprising that phase mixing produces amplification of the parallel electric field that accompanies
the Alfven wave in a collisionless plasma, although this
cannot be understood within the framework of ideal MHD theory which assumes $E_\parallel=0$.

Previous numerical studies of phase mixing
in a collisionless plasma have identified its implication in the generation of a parallel electric field and acceleration of electrons
\citep{Tsiklauri2005b,Tsiklauri2005,Tsiklauri2008}, see also \citep{Genot1,Genot2} in the magnetospheric context.
As stated above, the same features were established already some time ago, by Hasegawa and Chen,
for resonant absorption. Here, we provide a detailed 
 discussion of the role played by phase mixing in 
 both parallel electric field amplification and enhanced electron Landau damping of AWs in a collisionless plasma.

The calculations are based on the drift-kinetic theory presented in Section II, which is valid in the limit of low-frequency
fluctuations with $\omega\ll \omega_{ci}$, $\omega_{ci}$ being the ion cyclotron frequency. Phase mixing
and enhanced electron Landau damping of AWs in a collisionless low-$\beta$ plasma are considered
in Section III. Parallel electric field amplification is analyzed in Section IV. Conclusions and discussions are provided in
in Section V.

\section{Kinetic properties of the Alfven wave in a warm collisionless plasma}

Our starting point is the linearized drift-kinetic equation for the electrons:
\begin{equation}
\partial_{t}f_{1}+v_{\parallel}\nabla_\parallel f_{1}-\frac{e}{m_{e}}E_{\parallel}\partial_{v_{\parallel}}f_{0}=0
\end{equation}
The latter is supplemented by Maxwell's equations. Faraday's law is
\begin{equation}
E_{\parallel}=-\nabla_{\parallel}\phi -\frac{1}{c}\frac{\partial A_{\parallel}}{\partial t},
\end{equation}
$\phi$ is the electric potential, $A_{\parallel}$ the parallel component of the
vector potential.  The parallel component of Ampere's law reads
\begin{equation}
\nabla^{2}_{\perp}A_{\parallel}=\frac{4 \pi e}{c}\int v_{\parallel}f_{1} dv_{\parallel}.
\end{equation}
The above system is closed by the quasi-neutrality condition, which in the limit $k_{\perp}\rho_{i}\ll 1$, reads
\begin{equation}
n_{0}\rho_{i}^{2}\nabla_{\perp}^{2}\frac{e\phi}{T_{0i}}=\int f_{1} dv_{\parallel},
\end{equation}
$\rho_{i}$ is the thermal ion Larmor radius at the temperature $T_{0i}$ and $n_{0}$ is the background density.
We set the Boltzmann constant to unity which means that the temperature has the unit of energy. 
While this so-called gyrokinetic Poisson equation [Eq.(4)] includes the effect associated with the perpendicular ion polarization drift,
the electron response along the perturbed field lines is described by the drift-kinetic equation [Eq.(1)].

We assume a small deviation $f_{1}$ from an equilibrium Maxwellian distribution $f_{0}$:
\begin{equation}
f_{0}(v_{\parallel})=\frac{n_{0}}{\sqrt{\pi}v_{te}}e^{-v_{\parallel}^{2}/v_{te}^{2}}.
\end{equation}
The electron density perturbation is defined as $n_{e}=\int f_{1}dv_{\parallel}$
and the parallel current perturbation as $J_{\parallel}=-e\int v_{\parallel}f_{1}dv_{\parallel}=-en_{0}u_{\parallel e}$,
$u_{\parallel e}$ being the electron parallel velocity, and the electron pressure perturbation is defined as $P_{e}=m_{e}\int v_{\parallel}^{2}f_{1}d v_{z}$
Hence, Ampere's law and Poisson law can be written respectively as $\nabla^{2}_{\perp}A_{\parallel}=-(4/\pi c)J_{\parallel}$
and $\rho_{i}^{2}\nabla_{\perp}^{2} e\phi/T_{0i}=n_{e}/n_{0}$. On one hand, taking the zeroth order moment of the electron kinetic equation provides the electron continuity equation:
\begin{equation}
\frac{\partial n_{e}}{\partial t}+n_{0}ik_{\parallel}u_{\parallel e}=0
\end{equation}
On the other hand, the first moment provides the parallel electron momentum equation :
\begin{equation}
n_{0}m_{e}\frac{\partial u_{\parallel e}}{\partial t}=-ik_{\parallel}P_{e}-n_{0}eE_{\parallel},
\end{equation}
It is usual to refer to the last equation as the Ohms's law and  $P_{e}=n_{e}T_{0e}$ for an isothermal plasma.
Therefore, there are two possible sources of parallel electric field associated with the electron dynamics
: inertia and pressure (or density) variations along the field lines.
The continuity equation combined with Poisson law, yields a vorticity equation :
\begin{equation}
\frac{\partial}{\partial t}\rho_{i}^{2}\nabla_{\perp}^{2}\frac{e\phi}{T_{0i}}+\frac{c}{4\pi e n_{0}}ik_{\parallel}\nabla_{\perp}^{2}A_{\parallel}=0.
\end{equation}
Neglecting first the effects of electron inertia and electron pressure gradient in Ohm's law yields the MHD Ohm's law $E_{\parallel}=0$, i.e.
\begin{equation}
\frac{1}{c}\frac{\partial A_{\parallel}}{\partial t}=-ik_{\parallel}\phi.
\end{equation}
Introducing the stream and flux function for the velocity $\mathbf{u}_{\perp}=\mathbf{z}\times \nabla_{\perp}\varphi$,
 and the magnetic field $\mathbf{B}_{\perp}/\sqrt{4\pi n_{0}m_{i}}=\mathbf{z}\times \nabla_{\perp}\psi$, defined as $\varphi=(c/B_{0})\phi$ and $\psi=-A_{\parallel}/\sqrt{4\pi n m_{i}}$, gives
\begin{equation}\label{r1}
\frac{\partial}{\partial t}\nabla_{\perp}^{2}\varphi=v_{A}ik_{\parallel}\nabla_{\perp}\psi,
\end{equation}
\begin{equation}\label{r2}
\frac {\partial \psi}{\partial t}=v_{A}ik_{\parallel}\varphi,
\end{equation}
with $v_{A}=B_{0}/\sqrt{4\pi n_{0} m_{i}}$ being the Alfven velocity.
These two equations are the standard linearized reduced-MHD equations describing shear-Alfven waves with frequency :
\begin{equation}
\omega=\pm v_{A}k_{\parallel}.
\end{equation}
In the case where the parallel electric field is produced by density fluctuation in Ohm's law, we have
$E_{\parallel}=-ik_{\parallel}T_{0e}(n_{e}/n_{0})$. Using the Poisson equation,
 $E_{\parallel}=-ik_{\parallel}\rho_{s}^{2}\nabla_{\perp}^{2} \phi$, which also reveals the vortical
 nature of the parallel electric field. The parameter $\rho_{s}=c_{s}/\omega_{ci}=\sqrt{T_{0e}/T_{0i}}\rho_{i}$ is
  the ion gyroradius at the
electron temperature. By including this parallel electric field in Ohm's law, an extension of the previous reduced-MHD system now takes the form
\begin{equation}\label{r1}
\frac{\partial}{\partial t}\nabla_{\perp}^{2}\varphi=v_{A}ik_{\parallel}\nabla_{\perp}\psi,
\end{equation}
\begin{equation}\label{r3}
\frac {\partial \psi}{\partial t}=v_{A}ik_{\parallel}(\varphi-\rho_{s}^{2}\nabla^{2}_{\perp}\varphi),
\end{equation}
which describes the dynamics of kinetic Alfven waves with frequency
\begin{equation}
\omega=\pm v_{A}k_{\parallel}\sqrt{1+\rho_{s}^{2}k_{\perp}^{2}}.
\end{equation}
It is worth noticing that equations (\ref{r1})-(\ref{r3}) can also be obtained directly from two-fluid MHD theory by retaining
the Hall and electron pressure effects in Ohm's law\citep{Bian2009}.  
Using the above results, it is easily seen that for kinetic Alfven waves, the magnitude of the parallel
electric field is related to $B_{\perp}$ by
\begin{equation}
E_{\parallel}=\frac{v_{A}}{c}k_{\parallel}\frac{k_{\perp}\rho_{s}^{2}}{\sqrt{1+k_{\perp}^{2}\rho_{s}^{2}}} B_{\perp}
\end{equation}

The above fluid derivation of the Alfven wave frequency gives the same result as its kinetic counterpart,
however the latter, which is presented below, is more complete in the sense that it also provides the imaginary
part associated with Landau damping.
The electron kinetic equation can be solved for the perturbed distribution function $f_{1}$, i.e.
\begin{equation}
f_{1}=i\frac{e}{m_{e}}E_{\parallel}\frac{2n_{0}}{\sqrt{\pi}k_{\parallel}v^{3}_{te}}\frac{v_{\parallel}/v_{te}}
{v_{\parallel}/v_{te}-\omega/k_{\parallel}v_{te}}e^{-v_{\parallel}^{2}/v_{te}^{2}}.
\end{equation}
Some notations are introduced : $x=v_{\parallel}/v_{te}$, $\alpha=\omega/k_{\parallel}v_{te}$ and
\begin{equation}
Z_{n}(\alpha)=\frac{1}{\sqrt{\pi}}\int \frac{x^{n}}{x-\alpha}e^{-x^{2}}dx,
\end{equation}
with $Z_{0}(\alpha)$ being the standard plasma dispersion function. We also summarize some properties
 of the functions $Z_{n}$: $Z_{1}=1+\alpha Z_{0}$, $Z_{2}=\alpha Z_{1}$.
Moreover, in the limit $\alpha\ll 1$
\begin{equation}
Z_{0}(\alpha)\sim -2\alpha+i\sqrt{\pi}(1-\alpha ^{2}).
\end{equation}
Using the above properties, it follows that the density and current perturbations are related to the parallel
electric field through:
\begin{equation}
\int f_{1}dv_{\parallel}=\frac{2ien_{0}}{m_{e}k_{\parallel}v_{te}^{2}}[1+\alpha Z_{0}(\alpha)]E_{\parallel},
\end{equation}
 for the density, and
\begin{equation}
\int f_{1}v_{\parallel}dv_{\parallel}=\frac{2ien_{0}\omega}{m_{e}k^{2}_{\parallel}v_{te}^{2}}[1+\alpha Z_{0}(\alpha)]E_{\parallel}.
\end{equation}
Hence, the relation between parallel current and parallel electric field is
\begin{equation}
J_{\parallel}=\frac{-i \omega}{4\pi k_{\parallel}^{2}\lambda_{De}^{2}}[1+\alpha Z_{0}(\alpha)] E_{\parallel}.
\end{equation}
It is convenient to define a collisionless plasma conductivity $\sigma$
as
\begin{equation}
J_{\parallel}=\sigma E_{\parallel}
\end{equation}
Its imaginary part results in the dispersion of the Alfven wave and the its real part yields the
 collisionless dissipation. In the limit $\alpha\equiv \omega/k_{\parallel}v_{te} \ll 1 $, the real part is
\begin{equation}
\sigma_{r}\simeq \frac{e^{2}m_{e}^{1/2}n_{0}\omega^{2}}{k_{\parallel}^{3}T_{0e}^{3/2}}
\end{equation}
This also gives the energy per united time transferred to the electrons
through the relation :
\begin{equation}
Q=Re(J_{\parallel}E^{*}_{\parallel})
\end{equation}
i.e.
\begin{equation}
Q=\frac{\sqrt{\pi}\omega^{2}}{k_{\parallel}^{3}\lambda_{De}^{2}v_{te}}U_{E_{\parallel}}
\end{equation}
with $\lambda_{De}$ being the electron Debye length and
$U_{E_{\parallel}}=\mid E^{2}_{\parallel}\mid/8 \pi$ being the energy density of the parallel component of the electric field.
It is in fact a standard result that the asymptotic $\omega t\gg1$ averaged power transferred to electrons,
$Q=\int v_{\parallel} <-eE_{\parallel}f_{1}>dv_{\parallel}$ due to the presence of a
 an harmonic electric field fluctuation $E_{\parallel}= \cos(k_\parallel z -\omega t)$ is
\begin{equation}
Q=-\pi \frac{e^{2}E^{2}_{\parallel}}{2m_{e}k_{\parallel}}[v_{\parallel}\frac{\partial f_{0}}{\partial v_{\parallel}}]_{v_{\parallel}=\omega/k_{\parallel}}.
\end{equation}
This can easily be verified from Eq.(1) and for a Maxwellian distribution it is equivalent to Eq.(26).
Using the relation between $E_{\parallel}$ and $B_{\perp}$, $Q$ can finally be expressed in term of the magnetic energy, $U_{B}=\mid B^{2}_{\perp}\mid/8 \pi$,
\begin{equation}
Q=\frac{\sqrt{\pi}\omega^{2}}{k_{\parallel}v_{te}}\frac{k_{\perp}^{2}\rho_{s}^{2}}{1+k_{\perp}^{2}\rho_{s}^{2}}U_{B}.
\end{equation}
The coefficient of proportionality between $Q$ and $U_{B}$, which has the dimension of the inverse of a time, is
nothing else than the Landau damping rate.

The Landau damping rate is now directly obtained, without any reference to its physical meaning, from the complex dispersion relation.
The kinetic dispersion relation is obtained from : $\nabla_{\perp}^{2}A_{\parallel}=i\omega/(k_{\parallel}^{2}\lambda_{De}^{2}c)[1+\alpha Z_{0}(\alpha)]E_{\parallel}$, $\rho_{s}^{2}\nabla_{\perp}^{2}\phi=i/k_{\parallel}[1+\alpha Z_{0}(\alpha)]E_{\parallel}$
and $E_{\parallel}=-ik_{\parallel}\phi +i\omega A_{\parallel}/c$.
It is
\begin{equation}
\rho_{s}^{2}k^{2}_{\perp}+(1-\frac{\omega^{2}}{k_{\parallel}^{2}v_{A}})[1+\alpha Z_{0}(\alpha)]=0.
\end{equation}
This is the general complex dispersion relation for the dispersive Alfven wave. In the limit $\alpha \ll 1$, it reads
\begin{equation}
\omega^{2}=k^{2}_{\parallel}v_{A}^{2}[1+k_{\perp}^{2}\rho^{2}_{s}(1-i\sqrt{\pi}\alpha)].
\end{equation}
Its real part corresponds to the frequency of the kinetic Alfven wave which was also derived from fluid
theory above. Its imaginary part, which corresponds to the Landau damping
rate [see also Eq.(28)] and reads
\begin{equation}
\gamma(\mathbf{k})=\frac{\sqrt{\pi}}{2}\frac{v^{2}_{A}}{v_{te}}k_{\parallel}k_{\perp}^{2}\rho_{s}^{2}.
\end{equation}
Most calculations above were finalized in the limit $\alpha \ll1$, in the opposite limit of $\alpha \gg 1$ one obtains the frequency and damping rate of the inertial Alfven wave, which has its parallel electric field balanced by the electron inertia in Ohm's law. For frequency
$\omega \sim k_{\parallel}v_{A}$, $\alpha\sim v_{A}/v_{te}$, hence the kinetic Alfven wave regime corresponds to
$v_{A}/v_{te}\ll 1$ and the inertial Alfven wave regime to $v_{A}/v_{te}\gg 1$. In the following we continue to focus
on the warm plasma regime corresponding to $1\gg \beta_{e}\gg m_{e}/m_{i}$.

\section{Phase Mixing}

Phase mixing of a shear Alfven wave packet can be considered in the framework of an eikonal description:
\begin{equation}
\frac{d\mathbf{x}}{dt}=\nabla_{k}\omega,
\end{equation}
\begin{equation}
\frac{d\mathbf{k}}{dt}=-\nabla_{x}\omega,
\end{equation}
with $\omega=\pm k_{\parallel}v_{A}$.
These are the characteristics of the wave-kinetic equation
\begin{equation}
\frac{\partial {e_{\pm}}}{\partial t}+\nabla_{k}\omega.\nabla_{x}e_{\pm}-\nabla_{x}\omega \nabla_{k}e_{\pm}=-\gamma(\mathbf{k}) e_{\pm}.
\end{equation}
In the latter equation $e_{\pm}$ are the amplitudes of the
wave-packets corresponding to $\omega=\pm k_{\parallel}v_{A}$ and $\gamma(\mathbf{k})$ is 
a wave-number dependent damping rate. Following the trajectory of a wave-packet in phase space $(\mathbf{x},\mathbf{k})$, its amplitude
evolves according to :
\begin{equation}
\frac{d e_{\pm}}{dt}=-\gamma(\mathbf{k}) e_{\pm}.
\end{equation}
The latter equation is integrated to give
\begin{equation}
e_{\pm}(t)=e_{\pm}(0)\exp (-\int \gamma(\mathbf{k})dt).
\end{equation}
The principle of phase-mixing is simple: for any damping rate $\gamma$ which is an increasing function of $k$, any mechanism producing an
increase in $k$ as a function of time results also in a smaller damping time scale. This is precisely the situation when the
Alfven wave packet propagates along field lines with a transverse variation of the Alfven speed: the wave packet is sheared. In this case,
say $\mathbf{v}_{A}(x)=-v_{A}'x\mathbf{z}$, $\mathbf{z}$ being the unit vector in the parallel direction and $x$ the transverse
coordinate, then
\begin{equation}
\frac{dk_{\perp}}{dt}=k_{\parallel}v_{A}',
\end{equation}
with by definition $
v_{A}'=v_{A}/L_{\perp}$, $L_{\perp}$ being the characteristic length of the transverse inhomogeneity and $k_{\parallel}=k_{\parallel}(t=0)$.
This means that $k_{\perp}$ increases linearly with time due to differential advection of the wave packets along the field lines, i.e.
\begin{equation}
k_{\perp}(t)=k_{\parallel}v_{A}'t,
\end{equation}
where we have taken $k_{\perp}(t=0)=0$ without loss of generality.

For a resistive MHD Ohms' law, $E_{\parallel}=\eta J_{\parallel}$ the following results are well known. The damping rate is
$\gamma(\mathbf{k})=\eta c(k_{\perp}^{2}+k_{\parallel}^{2})/4\pi=D_{m}(k_{\perp}^{2}+k_{\parallel}^{2})$, this is the fourier transform of the operator responsible
for magnetic diffusion in the induction equation. Hence
\begin{equation}
e_{\pm}(t)=e_{\pm}(0)\exp[-D_{m}k^{2}_{\parallel} \int(1+v_{A}'^{2}t^{2})dt],
\end{equation}
which in the limit $t\gg v_{A}'^{-1}$ yields
\begin{equation}
e_{\pm}(z)\sim e_{\pm}(0)\exp(-\frac{D_{m}v_{A}'^{2}k^{2}_{\parallel}}{3}t^{3})
\end{equation}
Since, $z=v_{A}t$, we also have
\begin{equation}
e_{\pm}(t)\sim e_{\pm}(0)\exp(-\frac{D_{m}v_{A}'^{2}\omega^{2}}{3v_{A}^{5}}z^{3}),
\end{equation}
for an Alfven wave excited at $z=0$ with frequency $\omega$.
In a collisionless plasma, when the dissipation is provided by electron Landau damping, with damping rate
$\gamma(\mathbf{k})=\sqrt{\pi}v^{2}_{A}k_{\parallel}k_{\perp}^{2}\rho_{s}^{2}/2v_{te}$, the equivalent expressions are:
\begin{equation}
e_{\pm}(t)=e_{\pm}(0)\exp(-\frac{\sqrt{\pi}}{6}\frac{v_{A}^{2}v_{A}'^{2}}{v_{te}}\rho_{s}^{2}k_\parallel^{3}t^{3}),
\end{equation}
and also
\begin{equation}
e_{\pm}(z)=e_{\pm}(0)\exp(-\frac{\sqrt{\pi}}{6}\frac{v_{A}'^{2}}{v_{A}^{4}v_{te}}\rho_{s}^{2}\omega^{3}z^{3}),
\end{equation}
for an Alfven wave excited at $z=0$ with frequency $\omega$. Hence, the phase mixing time scale is
\begin{equation}
\tau_{pm}\sim \frac{v_{te}^{1/3}L_{\perp}^{2/3}}{v_{A}^{4/3}\rho_{s}^{2/3}k_{\parallel}},
\end{equation}
and the phase mixing length scale is
\begin{equation}
l_{pm}\sim \frac{v_{A}^{2/3}v_{te}^{1/3}L_{\perp}^{2/3}}{\rho_{s}^{2/3}\omega}.
\end{equation}
Notice that the scaling of the phase mixing length scale with the frequency $\omega$
in the spatial problem is different from that of resistive MHD phase mixing since
the collisionless conductivity associated with electron Landau damping depends on $\omega$, contrary to Spitzer conductivity.
However, the dependence with time or distance of the decay law, like $\exp (-\alpha_{1} t^{3})$ or $ \exp (-\alpha_{2}z^{3})$ are
 similar to resistive MHD phase mixing. The physical reason is obviously the common scaling of
 the damping rate $\gamma(\mathbf{k})$ with $k_{\perp}$
in the collisional and collisionless case.

The following comments are due. The enhanced electron Landau damping associated with phase-mixing was first
considered by \citep{voit1}. They derived a relation identical to Eq.(45) [see equations (30) and (11) in \citep{voit1}]. 
Moreover, results of the Particles-In-Cell (PIC) simulations carried by \citep{Tsiklauri2008} have produced
 $l_{pm}\propto \omega^{-\zeta}$ with $\zeta\simeq 1.10$, for the dependence of the phase mixing length scale $l_{pm}$
 with frequency $\omega$. They also report that the parallel electric field associated with the Alfven wave is primarily balanced by the electron pressure gradient in their simulations. They attribute the scaling of $l_{pm}$ with $\omega$ to the effect of
 an "anomalous resistivity" due to "scattering of particles by magnetic fields" which "plays an effective
 role of collisions". Here, we emphasize that their PIC simulation results can be accurately interpreted as the "normal" effect
  of electron Landau damping of the KAW
 since it gives $l_{pm}\propto \omega^{-\zeta}$ with $\zeta= 1$. 
 We now elaborate on the parallel electric field amplification which is observed in the simulations.

\section{Parallel electric field generation}

For an Alfven wave created by a source through perturbation of the background magnetic field, a parallel
electric field is produced, provided $k_{\perp}$ is finite, which is given by Eq.(16):
\begin{equation}
E_{\parallel}=\frac{v_{A}}{c}k_{\parallel}\frac{k_{\perp}\rho_{s}^{2}}{\sqrt{1+k_{\perp}^{2}\rho_{s}^{2}}} B_{\perp}
\end{equation}
It is this parallel electric field which is responsible for the Landau damping of the wave (see above).
For a given $k_{\parallel}$ and $\delta B_{\perp}$, this electric field is amplified provided that the $k_{\perp}$ associated
with the wave field is also amplified, $E_{\parallel}$ being a monotonic increasing function of $k_{\perp}$. However,
$E_{\parallel}(k_{\perp})$ also reaches a plateau for $k_{\perp}\rho_{s}\sim 1$, which is the boundary between the MHD and the dispersive
regime. Indeed,
\begin{equation}
E_{\parallel}=\frac{v_{A}}{c}k_{\parallel}k_{\perp}\rho_{s}^{2}B_{\perp}
\end{equation}
for $k_{\perp}\rho_{s}\ll 1$ and $E_{\parallel}$ reaches its maximum, of the order of
\begin{equation}
E_{\parallel}=\frac{v_{A}}{c}k_{\parallel}\rho_{s}B_{\perp}
\end{equation}
when $k_{\perp}\rho_{s}\sim 1$ or larger. Therefore, significant amplification of this parallel electric field can only occur in the range of wave-numbers where the wave is non-dispersive, i.e. it behaves as a shear-Alfven wave with frequency $\omega\simeq\pm k_{\parallel}v_{A}$
and, hence, it can be subject to standard phase mixing.

From the results of the previous section we obtain the dependence with time of the parallel electric field strength
during the phase mixing process :
\begin{equation}
\widetilde{E}_{\parallel}(t)=v_{A}'k^{2}_{\parallel}\rho_{s}^{2}t
\exp(-\frac{\sqrt{\pi}}{6}\frac{v_{A}^{2}v_{A}'^{2}}{v_{te}}\rho_{s}^{2}k_\parallel^{3}t^{3}),
\end{equation}
where a normalized electric field $\widetilde{E}_{\parallel}=E_{\parallel}/(B_{\perp}(0)v_{A}/c)$
has been defined. The variation with time $\widetilde{E}(t)$ has the form $\beta_{1} t \exp(-\alpha_{1}t^{3})$, with a growth phase
followed by a decay phase typical of the alternating field aligned current during phase mixing. Since $z=v_{A}t$, then
\begin{equation}
\widetilde{E}_{\parallel}(z)=\frac{v_{A}'\omega^{2}\rho_{s}^{2}z}{v_{A}^{3}}
\exp(-\frac{\sqrt{\pi}}{6}\frac{v_{A}'^{2}}{v_{A}^{4}v_{te}}\rho_{s}^{2}\omega^{3}z^{3}),
\end{equation}
which has the form $\beta_{2}z\exp(-\alpha_{2}z^{3})$, for an Alfven wave excited at $z=0$ with frequency $\omega$. The
above defined phase mixing time/length scales are precisely the scales associated with the amplification of the parallel electric field, i.e. the time/length scales
for the parallel electric field to reach its maximum value given by :
\begin{equation}
\widetilde{E}\sim \frac{\omega v_{te}^{1/3}\rho^{4/3}_{s}}{v_{A}^{4/3}L^{1/3}}
\end{equation}
with $\omega\simeq k_{\parallel}v_{A}$
\section{Conclusions}
Previous PIC (Particles-In-Cell) simulations of "collisonless phase mixing" of Alfven waves \citep{Tsiklauri2005b,Tsiklauri2005,Tsiklauri2008}
have identified its
relation to the generation of a parallel
electric field and acceleration of electrons. Importance of this parallel
electric field was first pointed out by Hasegawa and Chen in the context
 of resonant absorption, who also showed that the dominant energy dissipation
 of the Alfven wave, in a collisionless low-$\beta$ plasma, involves energy transfer to the electrons\citep{Hasegawa1976a}.
The role of electron Landau damping in "collisonless phase mixing" was also considered by \citep{voit1,voit2}
 
Focusing on the "kinetic" regime of the dispersive Alfven wave, when $v_{A}/v_{te}\ll 1$, we provided
a detailed discussion of the role played by phase mixing in both parallel electric field amplification and enhanced electron Landau damping
of the wave. 

Qualitatively, the physics of collisionless phase mixing can be summarized as follow. A parallel electric field accompanies
the propagation of Alfven waves with finite $k_{\perp}$ because they compress the plasma. The magnitude of this electric field
is an increasing function of $k_{\perp}$ that saturates in the dispersive range when $k_{\perp}\rho_{s}\sim 1$ or larger.
Therefore, any mechanism that produces an increase in $k_{\perp}$ also
leads to the amplification of the parallel electric field associated with the Alfven wave.
Phase mixing is such a mechanism, independently of the energy dissipation
channel. Phase mixing is a special occurrence of energy-conserving cascade\citep{Bian2008}. Such
 a cascade, predominantly involving perpendicular wave-numbers, is generally attributed to non-linear interactions, i.e.
 to turbulence.
 Existence of this parallel electric field
and the dependence of its magnitude with $k_{\perp}$ yield a Landau damping rate which scales like $k_{\perp}^{2}$, just as
visco-resistive damping. This can be demonstrated very simply in the framework of drift-kinetic theory. Therefore, in a collisionless plasma, phase mixing leads to enhanced electron Landau damping of the Alfven wave
in a manner which is very similar to the well-studied case of enhanced visco-resistive damping.
As a consequence, once the wave has Landau damped in a collisionless low-$\beta$ plasma, its energy has been transferred to the electrons, and the time and length
scales involved have been evaluated for small amplitude perturbations. Moreover, we studied the evolution
of the magnitude of the parallel electric field in the course of its amplification and calculated its maximum
value.

We argued that the scaling of the phase mixing length scale
 with frequency, $l_{pm}\propto \omega^{-\zeta}$ and $\zeta\simeq 1$, reported by \citep{Tsiklauri2008}
has a simple interpretation in term of electron Landau damping. PIC simulations of collisionless phase mixing
are valuable tools because they can provide
direct information on the modification of the electron distribution function
involved in the acceleration process,
see \citep{Tsiklauri2005b,Tsiklauri2005}, a feature which the present
kind of analysis is not capable of. For this, a theoretical framework is needed, e.g. quasi-linear theory. This is the subject of
ongoing work.

\begin{acknowledgements}

This work is supported by a STFC rolling grant (NHB, EPK) and an STFC Advanced
Fellowship (EPK). Financial support by the Leverhulme Trust
grant (F/00179/AY) and by the European Commission through the SOLAIRE
Network (MTRN-CT-2006-035484) is gratefully acknowledged.

\end{acknowledgements}

\bibliographystyle{aa}
\bibliography{refs2}

\end{document}